# A new method to detect the vortex glass phase and its evidence in YBCO

**M G Adesso, M Polichetti and S Pace**

Physics Department, SUPERMAT, INFM, University of Salerno, Via S. Allende, 84081, Baronissi (Salerno) Italy.

Corresponding author's e-mail: adesso@sa.infn.it

**Abstract.** The Vortex Glass phase has been unequivocally identified by analyzing the non linear magnetic response of type II superconductors. The method here introduced, more effective than the study of direct transport measurements, is based on a combined frequency dependence analysis of the real and imaginary part of the 1$^{st}$ and 3$^{rd}$ harmonics of the AC magnetic susceptibility. The analysis has been performed by taking into account both the components and the Cole-Cole plots (i.e. the imaginary part as a function of the real part). Numerical simulations have been used to individuate the fingerprints of the magnetic behaviour in the Vortex Glass phase. These characteristics allowed to distinguish the Vortex Glass phase from the other disordered phases, even those showing similar electrical properties. Finally, this method has been successfully applied to detect the Vortex Glass Phase in an YBCO bulk melt-textured sample.

## 1. Introduction

Experimental and theoretical studies about the "effective zero resistance" in the voltage-current characteristics of type-II superconductors, are widely under investigation in literature [1]. D. A. Huse et al. [2] argued the existence of a second order thermodynamic phase transition between a vortex phase, with a small but nonzero resistivity, and a truly superconducting phase, named *vortex glass*, with no mobile vortices and thus strictly zero resistivity. In this latter phase, the vortices were supposed frozen into a configuration determined by the competition between the interactions of the vortices with each other and with the microscopic impurities in the material [2]. The Kim-Anderson model [3] is generally used to describe the vortex dynamics, and in particular the thermal activation processes (Flux Creep [4]) in the phase with nonzero resistance, whereas several models [5-7], e.g. the Vortex Glass Collective Creep models [8], have been developed to explain the resistivity approaching to zero, characteristic of the Vortex Glass phase. These two different approaches are mainly distinguished for the dependence of the pinning potential ($U_p$) on the current density ($J$): a linear $U_p(J)$ is supposed in the Kim-Anderson model whereas all the vortex glass models are characterized by a non linear $U_p(J)$. Consequently, in presence of a vortex glass phase, a negative curvature has to be detected in a log$I$-log$V$ plot [2], for temperature $T$ lower than the vortex-glass phase transition temperature, $T_g$, in sharp contrast with the Kim-Anderson Flux Creep prediction, which generates a positive curvature [3, 4].

In literature there is an open question about the interpretation of the glass like behaviour of the vortex matter. Some experimental evidences can be found about this phase transition. Koch. et al. [9] showed a negative curvature in a temperature range $T < T_g$ in the Voltage-Current characteristics measured on epitaxial YBCO film samples, which can be interpreted in terms of the Vortex Glass phase. They also measured the critical exponents which are consistent with those expected for this phase transition [5, 9]. Nevertheless, the interpretation of the Koch experimental data [9] is controversial: Coppersmith et al. [10] and Landau et al. [11, 12] showed that the standard Kim-Anderson approach, modified with a inhomogeneous pinning potential, $U_p(x)$, depending on the spatial position $x$, also reproduced the qualitative features of the Koch's V-I characteristics [9], and in particular the negative curvature in the E-J curves at low temperatures, thus leaving still open the real interpretation of the Koch's data [12]. In this sense, the voltage-current characteristics are not suitable to clearly identify the presence of a vortex glass phase.

Here we show an innovative method to individuate the Vortex Glass phase, based on the analysis of the non linear magnetic response of the samples, in particular on the frequency dependence of the fundamental and higher harmonics of the AC magnetic susceptibility. Moreover, by a comparison between numerical results and experimental data, obtained on a YBCO melt-textured sample, we report the detection of the Vortex Glass phase in the analyzed material.

## 2. Simulations of harmonics of the AC magnetic susceptibility

The 1$^{st}$ and 3$^{rd}$ harmonics of the AC magnetic susceptibility as a function of the temperature have been simulated by integrating the one-dimensional non-linear diffusion equation ($x$ = linear dimension, $t$ = time) for the magnetic field inside the sample [13]:

$$\frac{\partial B}{\partial t} = \frac{\partial}{\partial x}\left[\left(\frac{\rho(B,J,T)}{\mu_0}\right) \cdot \left(\frac{\partial B}{\partial x}\right)\right] \quad (1)$$

where $\rho(B,J,T)$ is the resistivity associated to the vortex movements. Several $\rho(B,J,T)$ dependences on the field ($B$), temperature ($T$) and current density ($J$) have been investigated, corresponding to different V-I characteristics [13-15]. Here we report the simulations obtained by using the formula [5, 16]:

$$\rho(B,J,T) = \rho_{FF}(B,T) \cdot e^{-\left(\frac{U_p(J,T)}{k_B T}\right)}, \quad (2)$$

In the (2), $\rho_{FF}(B,T)$ is the Flux Flow resistivity, given by the Bardeen Stephen model [1, 17]:

$$\rho_{FF}(B,T) = \rho_n(T)\frac{B}{H_{c2}(T)}, \quad (3)$$

where $\rho_n$ is resistivity in the normal state (in the present simulation: $\rho_n(T) = 40 + 2.2 \cdot 10^{-1}[T + 273.16K]$) and $H_{c2}(T)$ is the temperature dependent upper critical field, chosen as in [13, 18]:

$$H_{c2}(T) = H_{c2}(0)\frac{\left[1-\left(\frac{T}{T_c}\right)^2\right]}{\left[1+\left(\frac{T}{T_c}\right)^2\right]}, \quad (4)$$

where $H_{c2}(0) = 112$ T, which is a typical value for YBCO bulk samples [13].

The $U_p(T,J)$ in the expression (2) is the pinning potential [1, 5], which depends on the temperature and the current density (in absence of the DC field, and supposing that the *ac* field dependence in $U_p$ is negligible):

$$U_p(T,J) = U_0 \cdot f(T) \cdot F(J,T) \qquad (5)$$

The current density dependence of $U_p$ is associated to the vortex dynamics and in particular to the Flux Creep models. In the simulations, the following density current dependence has been used [8]:

$$F(J,T) = \frac{1}{\mu}\left[\left(\frac{J_c(T)}{J}\right)^{\mu} - 1\right], \qquad (6)$$

with different $\mu$ values: 1/7 for single vortex, 3/2 for small bundle, 7/9 for large bundle Vortex Glass Collective Flux Creep [8] and $-1$ corresponding to the linear Kim-Anderson Flux Creep model [3]. $J_c$ in the (6) is the critical current density [1, 5].

On the other side, the temperature dependence of $U_p$ and $J_c$ is related to the flux pinning model. The behaviour of the harmonics with different pinning models has been already investigated [13]. Here we only report the results obtained by choosing the $\delta l$-type collective pinning [19, 5]. This choice is justified by the previously reported [20] strong evidence for the importance of $\delta l$ pinning in stoichiometric yttrium-based superconductors. Moreover, an experimental confirmation of the validity of this pinning model was also previously obtained in a polycrystalline YBCO sample obtained by the same batch as the one measured in this work [21]. For these reasons, the following temperature dependencies were used in the numerical simulations here shown:

$$U_p(T) \propto \left[1 - \left(\frac{T}{T_c}\right)^4\right] \equiv f(T) \qquad (7)$$

$$J_c(T) = J_c(0) \frac{\left[1 - \left(\frac{T}{T_c}\right)^2\right]^{5/2}}{\left[1 + \left(\frac{T}{T_c}\right)^2\right]^{1/2}} \qquad (8)$$

Moreover, $U_0$ and $J_c(0)$ are the pinning potential and the critical current density at $T = 0$, $H = 0$, $J = 0$, respectively. The values of the parameters used in the simulations are: $T_c = 91.6$ K, $J_c(0) = 10^9$ A·m, and $U_0/k_B = 1.6 \times 10^4$ K. These parameters have been experimentally obtained on the sample analysed in the present work. In particular, $T_c$ and $J_c(0)$ have been measured by magnetization curves, whereas $U_0/k_B$ has been obtained by a combined analysis of the 1st and 3rd harmonics of the AC magnetic susceptibility [22].

We performed simulations at a fixed amplitude of the AC magnetic field ($h_{AC} = 4$ Oe), without a DC field, at various AC frequencies ($\nu = 10.7, 107, 1070$ Hz), by choosing the above mentioned different Flux Creep models.

## 3. Numerical results

In this section, we summarize the comparison between all the numerically obtained results, in order to individuate the fingerprints of the Vortex Glass Phase.

*3.1. The Vortex Glass models vs the homogeneous Kim-Anderson case*

A preliminary analysis based on numerical simulations was already previously performed [23]. Here we also report the main obtained results, in order to clearly individuate the characteristics of the Vortex Glass phase with respect to the other disordered phases. In figure 1, the temperature dependence of the imaginary part of the 1$^{st}$ harmonics, at various frequencies of the AC magnetic field, is shown, as simulated in the Vortex Glass Collective Creep model, in the Small Bundle regime. Similar results have been also obtained for Single Vortex [23] and Large Bundle regimes (not reported). The same conditions are used to simulate the 1$^{st}$ harmonics in the framework of the Kim-Anderson model, as reported in figure 2.

From figure 1, we can observe that, in the Vortex Glass phase, for increasing frequencies, the temperature of the peak in the imaginary part of the 1$^{st}$ harmonics, $T_p$, shifts towards higher temperatures and the height of the peak, $\chi_1^{"}(T_p)$, grows. Nevertheless, in the Kim-Anderson model (figure 2), $T_p$ also shifts towards higher temperatures if $\nu$ is increased, but $\chi_1^{"}(T_p)$ decreases.

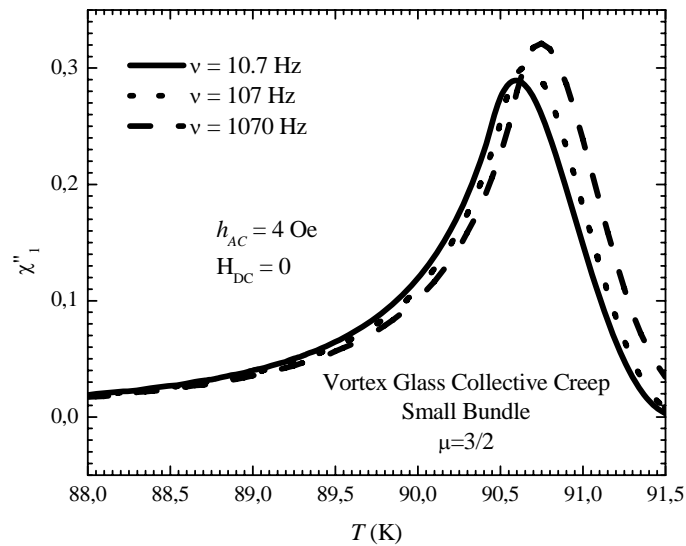

**Figure 1.** Imaginary part of the 1$^{st}$ harmonics as a function of the temperature as simulated by using the Vortex Glass Collective Creep in the Small Bundle regime, at various frequencies. The curves obtained for the Single Vortex and Large bundle regimes (not reported) show similar qualitative behaviours.

On the contrary, no qualitative differences can be observed in the behaviour of the real part of the 1$^{st}$ harmonics at various frequencies, in both the models [23]. An opposite behaviour with the frequency in the Kim-Anderson Creep and the Vortex Glass models can be also detected if we analyse the 1$^{st}$ harmonics Cole-Cole plots, as shown in figure 3.

In fact, the height of the maximum in the 1st harmonics Cole-Cole plots decreases for increasing frequencies in the Kim-Anderson model, whereas an opposite behaviour characterizes the Vortex Glass models, in all the regimes.

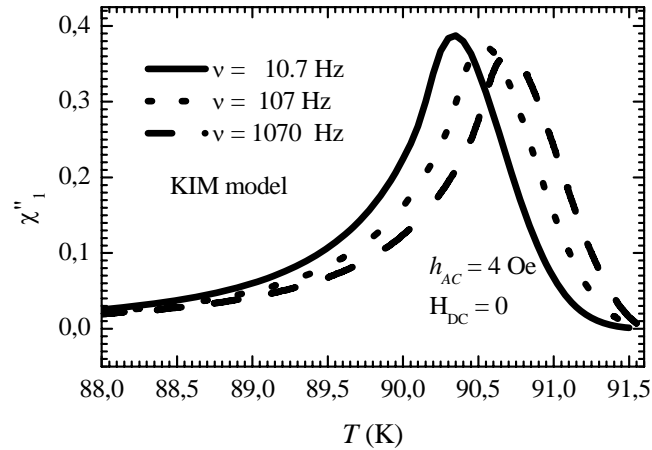

**Figure 2.** Temperature dependence of the imaginary part of the 1st harmonics at various frequencies, simulated by using the Kim-Anderson model.

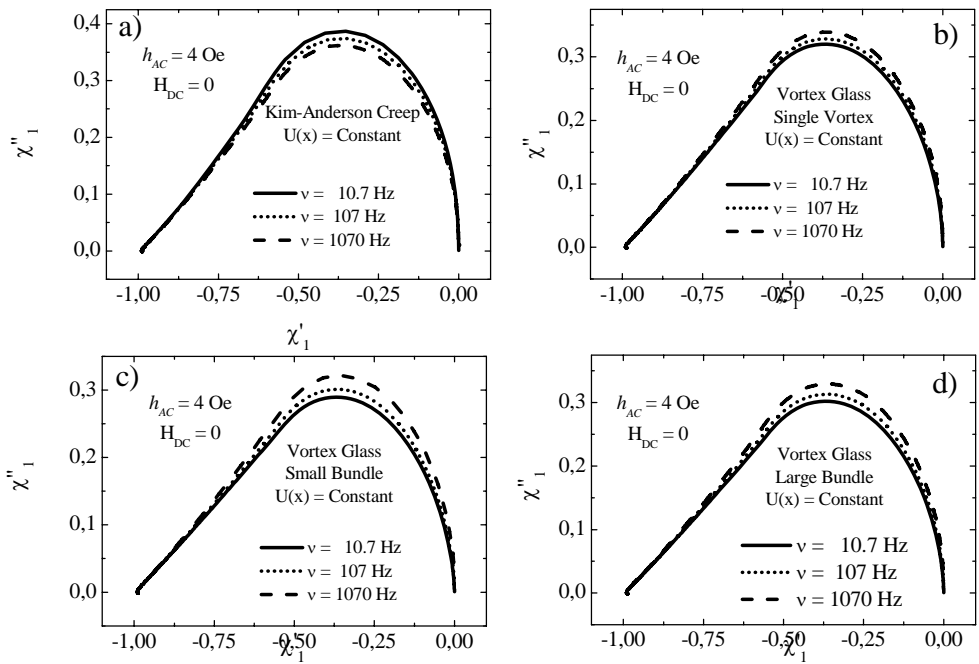

**Figure 3.** 1st harmonics Cole-Cole plots at various frequencies, simulated by using the Kim-Anderson model (a) and Vortex Glass Collective Creep model, respectively in the Single Vortex (b), Small Bundle (c) and Large Bundle regime (d).

A similar analysis can be also performed on the higher harmonics of the AC magnetic susceptibility. The main differences between the Vortex Glass and the "resistive" phase can be evidenced if we observe the real part, $\chi'_3(T)$, of the 3$^{rd}$ harmonics, whereas a similar behaviour has been detected in the imaginary part of the 3$^{rd}$ harmonics [23].

In figure 4, the real part of the 3$^{rd}$ harmonics of the AC susceptibility simulated in the Vortex Glass phase (in the Small Bundle regime) are shown, at various frequencies. In figure 5 the corresponding curves, as obtained by using the Kim-Anderson model, are reported.

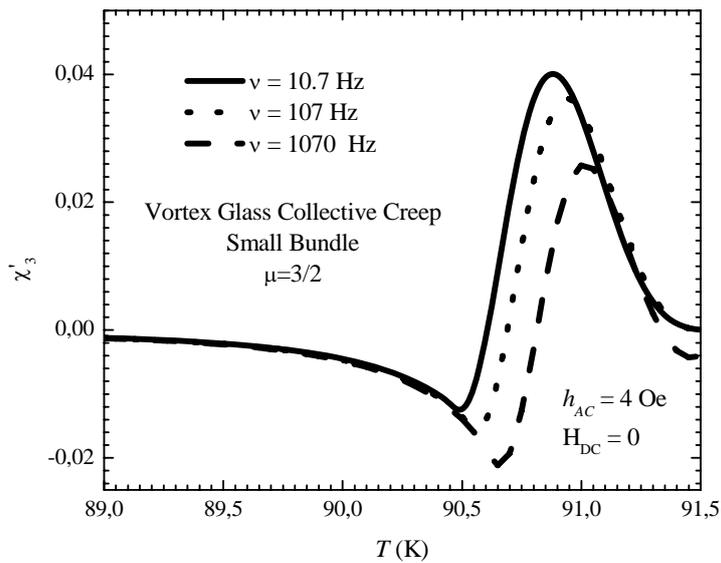

**Figure 4.** Temperature dependence of the real part of the 3$^{rd}$ harmonics at various frequencies, as simulated with the Vortex Glass Collective Creep model in the Small Bundle regime.

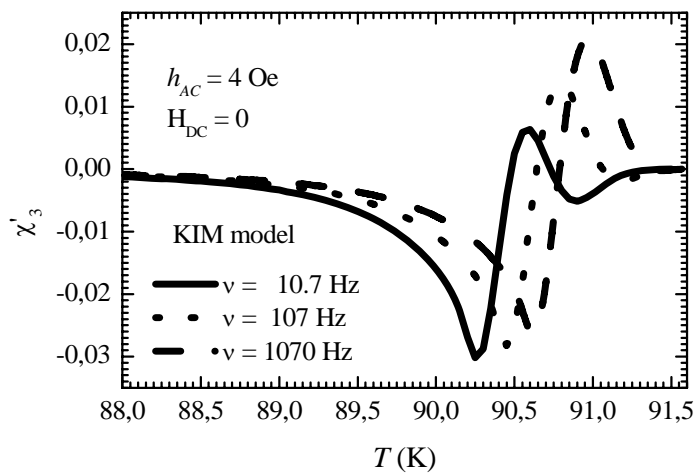

**Figure 5.** Temperature dependence of real part of the 3$^{rd}$ harmonics at various frequencies, as obtained by using the Kim-Anderson model.

From figure 4 and figure 5, we observe that, in all the considered models, $\chi_3'(T)$ shows a minimum and a maximum, both depending on the frequency. Nevertheless, for increasing frequencies, in the glass phase the absolute value of the minimum grows and the maximum decreases, whereas the behaviour is opposite in the Kim-Anderson framework.

A more distinct frequency response can be evidenced from the 3$^{rd}$ harmonics Cole-Cole plots, reported in figure 6 for all the considered models.

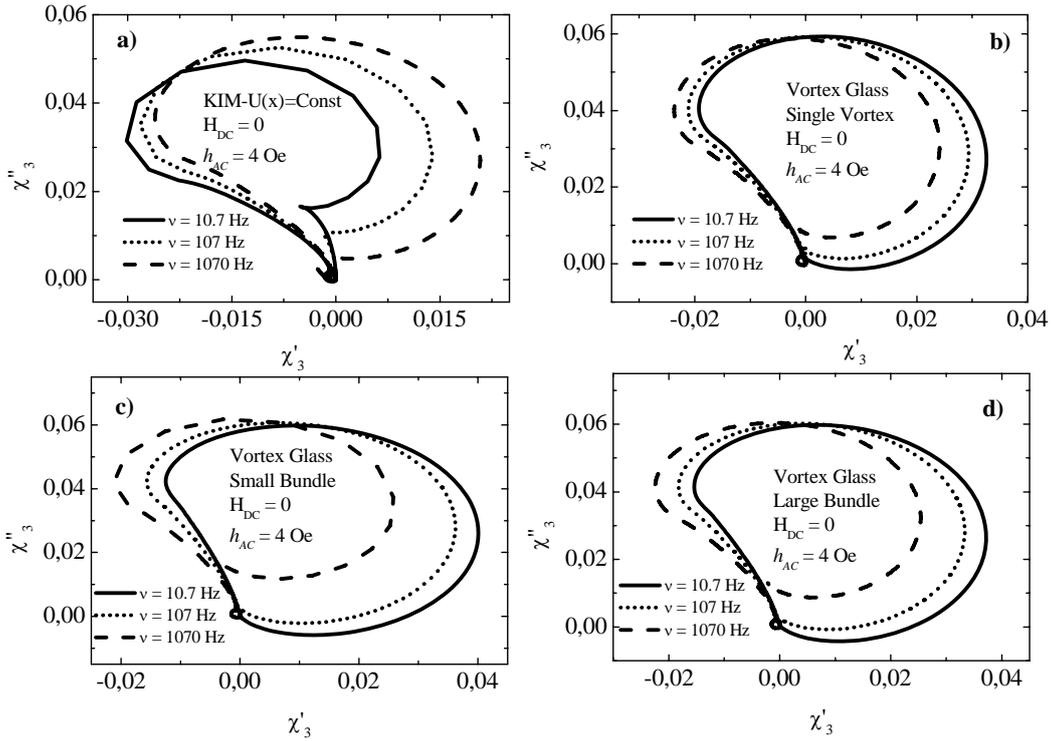

**Figure 6.** 3$^{rd}$ harmonics Cole-Cole plots at various frequencies, simulated by using a) the Kim-Anderson model and b)-d) the vortex glass collective creep model, respectively in the Single Vortex (b), Small Bundle (c) and Large Bundle (d) regime.

From figure 6 we observe that the 3$^{rd}$ harmonics Cole-Cole plots point towards the right semi-plane more and more for increasing frequencies in the Kim-Anderson model, whereas they tend to the left semi-plane in the Vortex Glass phase, for all the regimes.

In conclusion, we can state that the frequency dependence of the imaginary part of the 1$^{st}$ harmonics and the real part of the 3$^{rd}$ harmonics, together with both the 1$^{st}$ and 3$^{rd}$ harmonics Cole-Cole plots, without a DC field, allow to distinguish a Vortex Glass phase from the phase described by the Kim-Anderson model.

*3.2. A comparison between the Vortex Glass and the inhomogeneous Kim-Anderson case*

In literature [11, 12], the standard Kim-Anderson approach, modified with an inhomogeneous pinning potential, *U(x)*, is supposed to be equivalent to the Vortex Glass Collective Creep, because it reproduces some qualitative features of the Voltage-Current characteristics, associated to the occurrence of a Vortex Glass phase [2].

The choice of the $U(x)$ dependence does not influence the main features of the analysis [12] but, in order to reproduce the Koch [9] *V-I* characteristics, it is also necessary to include a further temperature dependence, strictly connected to the *x*-dependence [12]. According to Ref. [12], we chose the following model for $U(x,T)$:

$$U(x,T) \propto \left( |x| - a\left[1 - \left(\frac{T}{T_c}\right)\right]^k \cdot x^2 \right) \equiv h(x,T) \qquad (9)$$

where x=0 corresponds to the semi-thickness of the sample.
The parameters used in our simulations are: $a = 1$ and $k = 1.5$ [12].
In figure 7 the imaginary part of the 1$^{st}$ harmonics and the real part of the 3$^{rd}$ harmonics as a function of the temperature are reported at various frequencies, as simulated with this inhomogeneous pinning model within the Kim-Anderson framework. The corresponding 1$^{st}$ and 3$^{rd}$ harmonics Cole-Cole plots are plotted in figure 8.

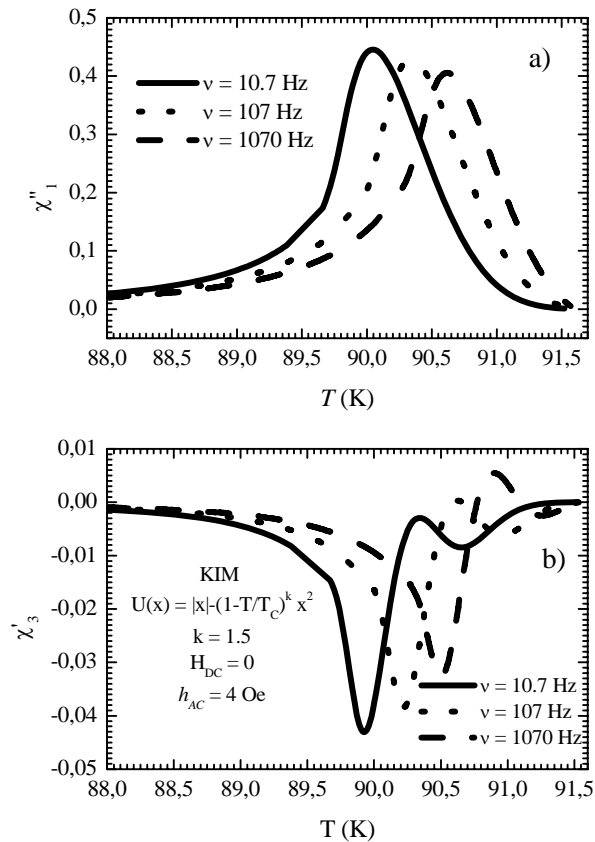

**Figure 7.** Imaginary part of the 1$^{st}$ harmonics (a) and real part of the 3$^{rd}$ one (b) as a function of the temperature, simulated by using an inhomogeneous Kim-Anderson model at various frequencies.

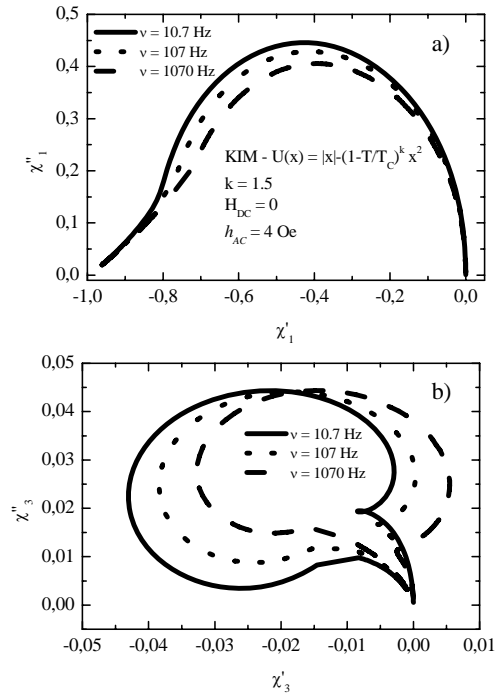

**Figure 8.** 1$^{st}$ (a) and 3$^{rd}$ (b) harmonics Cole-Cole plots at various frequencies, computed by using an inhomogeneous Kim-Anderson model.

It is possible to observe that the imaginary part of the 1$^{st}$ harmonics is characterized by a positive peak shifting towards higher temperatures with a decreasing height, for increasing frequencies. Moreover, two peaks can be distinguished in the real part of the 3$^{rd}$ harmonics: for increasing frequencies the height of the positive peak near $T_c$ increases, whereas the height of the negative peak at lower temperatures decreases in absolute value. The 1$^{st}$ harmonics Cole-Cole plots are characterized by a maximum which decreases for increasing frequencies, whereas the 3$^{rd}$ harmonics Cole-Cole plots tend to the right semi-plane for higher frequencies. All these dependences are similar to the results obtained with the standard Kim-Anderson model, and are opposite to the curves simulated with the Vortex Glass models.

Therefore, from this analysis, it is possible to conclude that the Vortex Glass phase has a different behaviour with respect to a phase described by the Kim-Anderson Flux Creep model, both in the homogeneous and inhomogeneous case.

### 4. A method to detect the Vortex Glass Phase

Summarizing the previous results, a good method to individuate a Vortex Glass phase is to analyse the behaviour of the harmonics of the AC magnetic susceptibility by changing the frequency of the AC magnetic field, without a DC field. In particular, a Vortex Glass Phase can be detected if, for increasing frequencies:

1) in the imaginary part of the 1$^{st}$ harmonics vs $T$, the height of the peak increases;
2) in the real part of the 3$^{rd}$ harmonics vs $T$, the absolute value of the minimum at low temperatures grows and the height of the maximum near $T_c$ decreases;
3) in the 1$^{st}$ harmonics Cole-Cole plots, the height of the maximum increases;
4) the 3$^{rd}$ harmonics Cole-Cole plots point towards the left semi-plane.

The harmonics described by the Kim-Anderson model have an opposite behaviour in all these cases. This analysis furnishes a method to experimentally detect a vortex glass phase by magnetic measurements, thus overcoming the problem connected to the "real zero resistivity" in the direct transport measurements.

## 5. Experimental evidence of the Vortex Glass Phase

The above introduced method is quite general and it could be applied to any type-II superconductors. Here we used it to analyze the experimental results obtained on an YBCO bulk melt textured sample. The measured sample was a melt grown YBCO sample, cut as an almost homogeneous slab (2mm x 3.1 mm x 4.8 mm), obtained by the same batch previously analysed [24]. An home made susceptometer has been used to measure the 1$^{st}$ and 3$^{rd}$ harmonics of the AC magnetic susceptibility as a function of the temperature at various frequencies ($\nu$=10.7, 107, 1607 Hz) and amplitudes ($h_{AC}$) of the AC magnetic field, both with and without an external DC field ($H_{DC}$). The harmonics have been measured by applying both the AC and the DC magnetic fields parallel to the longitudinal axis of the sample.

In figure 9, the imaginary part of the 1$^{st}$ harmonics and the real part of the 3$^{rd}$ one are plotted at various frequencies as a function of the temperature. In figure 10, the 1$^{st}$ and 3$^{rd}$ harmonics Cole-Cole plots are shown, as measured at the same conditions. In figure 11, the imaginary part of the 1$^{st}$ harmonics and the 1$^{st}$ harmonics Cole-Cole plots are shown, as measured at various amplitudes of the AC field, at a fixed frequency and without a DC field.

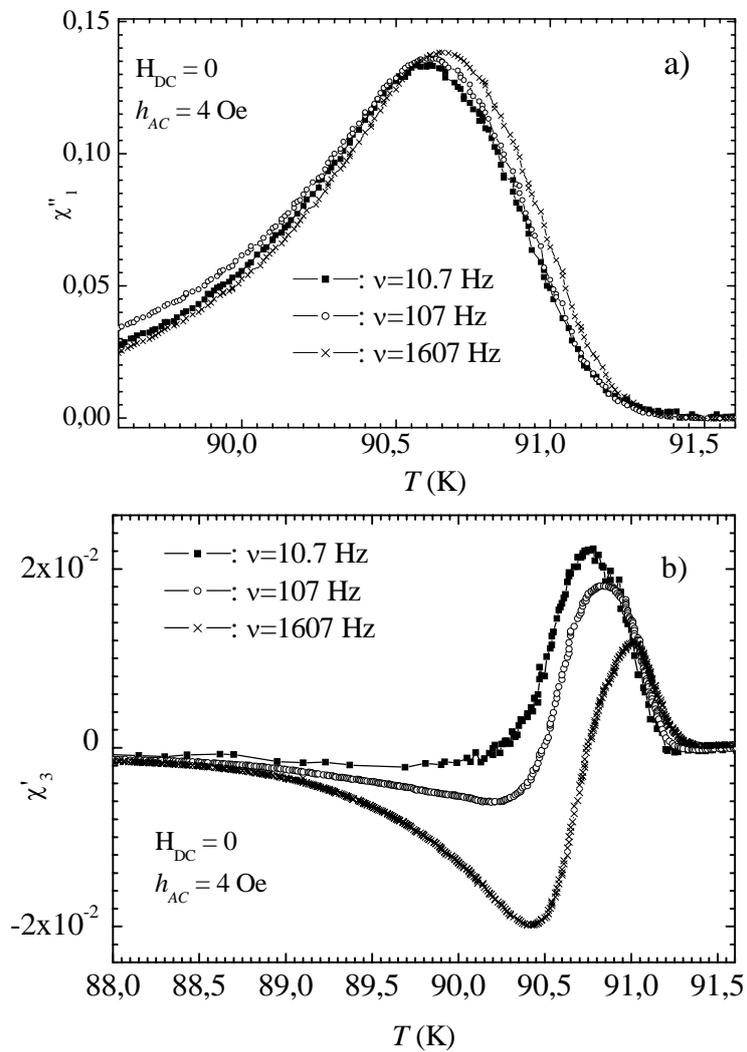

**Figure 9.** Imaginary part of the 1st harmonics (a) and real part of the 3rd harmonics (b) as a function of the temperature, as measured on an YBCO sample at various frequencies and at a fixed amplitude of the AC magnetic field.

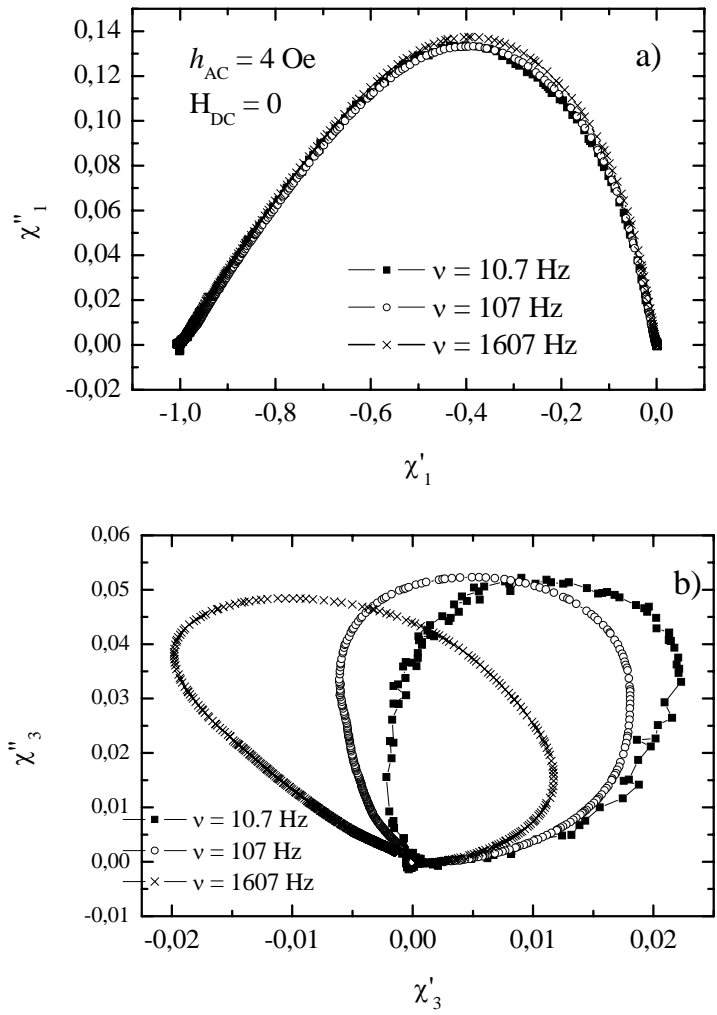

**Figure 10.** 1st (a) and 3rd (b) harmonics Cole-Cole plots, as measured on the YBCO sample at various frequencies.

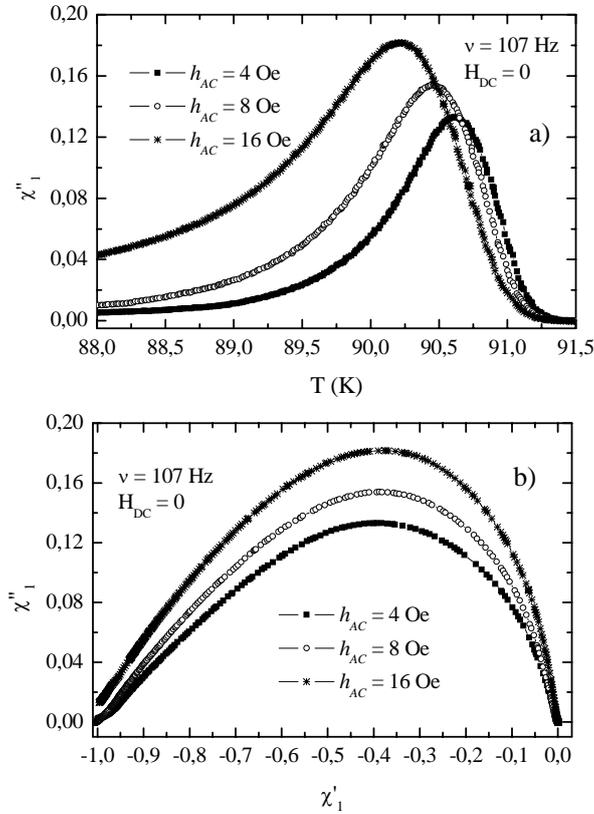

**Figure 11.** Imaginary part of the 1$^{st}$ harmonics (a) and 1$^{st}$ harmonics Cole-Cole plots (b) measured on YBCO melt-textured at various amplitude of the AC magnetic field, without a DC field.

The data reported in figures 9 – 11 suggest that the effects due to an eventual granularity, which could exist in our sample, are not present. In fact, as it is known [1, 28, 29], in the case of a granular sample, the AC magnetic response is characterized by two contributions, i.e. the inter-grain and the intra-grain components. These can be evidenced as a double peak in the imaginary part of the 1$^{st}$ harmonics vs Temperature curves (as well as a corresponding double-step transition in $\chi_1'(T)$ ), and as two joined dome-shaped curves in the 1$^{st}$ harmonics Cole-Cole plots. From figures 9-11, we deduced that the effects of granularity can be neglected, even if external parameters like the amplitude and the frequency of the AC magnetic fields are changed. The absence of the granularity has been also confirmed through measurements in presence of different DC fields (not shown here).

From figures 9 and 10 we observe that all the characteristics of the Vortex Glass phase, described in the previous section, have been detected. In particular, in the 3$^{rd}$ harmonics for increasing frequencies the absolute value of the minimum at low temperatures grows, the maximum near $T_c$ decreases, and the Cole-Cole plots tend toward the left semi-plane. Moreover, the presence of the Vortex Glass phase is also confirmed by a slightly, but still evident, growing of the maximum in $\chi_1'(T)$ and of the maximum in the 1$^{st}$ harmonics Cole-Cole plots with the frequency.

It is worth underlining that this behaviour in the 3$^{rd}$ harmonics Cole-Cole plots could be also interpreted in terms of edge barrier. In order to exclude this possibility, we also performed some measurements of harmonics at various DC fields. In figure 12 the 3$^{rd}$ harmonics Cole-Cole plots are reported, at various DC fields up to 400 Oe, as measured at a fixed AC field. The comparison between these curves and the simulations reported in Ref. [25] suggest that the surface effects can be disregarded. In fact, our curves have a lent / half cardioid shape, lie in a single quadrant, and behave almost independently of the applied DC field. All these characteristics are common to the curves simulated in [25] in case of a bulk pinning, and are totally different from those simulated in presence of edge effects that are cardioid in both the semi-planes.

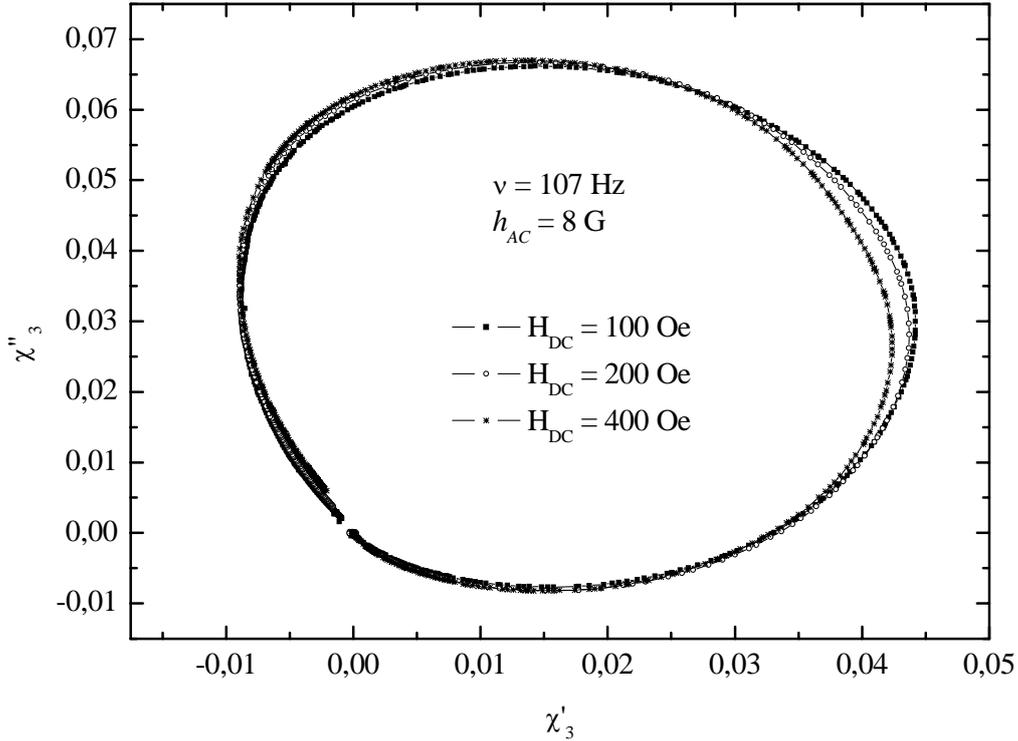

**Figure 12.** 3[rd] harmonics Cole-Cole plots, measured at a fixed frequency and amplitude of the AC magnetic field, at various DC fields, measured on the same YBCO sample: the half cardioid/lent shaped curves are in agreement with the simulations [25] obtained by including the bulk screening.

On the basis of these analysis and by using the method introduced in the previous section, finally we conclude that the data here reported in figures 9 and 10 are an experimental evidence of the occurrence of a Vortex Glass phase in a YBCO melt-textured sample, with negligible granularity and surface effects.

## 6. Conclusions

We introduced an effective method to identify the vortex glass phase in a type-II superconductor, based on the non linear magnetic response of the samples. In order to develop this method, numerical simulations of the harmonics of the AC magnetic susceptibility have been performed, at various frequencies of the AC magnetic field, by using different flux creep and flux pinning models. In particular, we demonstrated that the frequency dependence in the magnetic response simulated by using the Kim-Anderson flux creep model (both considering the homogeneous and the inhomogeneous flux pinning model) is different from the magnetic behaviour in the vortex glass phase. Moreover, thanks to a comparison between numerical results and experimental data, the Vortex Glass phase has been successfully detected in an YBCO bulk sample.